# DVS: Blood cancer detection using novel CNN-based ensemble approach

Dr. Md Taimur Ahad
Associate Professor
Department of CSE
Daffodil International Univarsity
Dhaka, Bangladesh
taimurahad.cse@diu.edu.bd

Israt Jahan Payel
Department of CSE
Daffodil International Univarsity
Dhaka, Bangladesh
israt15-3138@diu.edu.bd

Bo Song
Lecturer (Industrial Automation)
Section
School of Engineering
University of Southern Queensland
Bo.Song@usq.edu.au

Yan Li
Professor (Computing)
School of Mathematics, Physics and Computing
Toowoomba Campus
University of Southern Queensland
Yan.Li@usq.edu.au

***Abstract* —** Blood cancer can only be diagnosed properly if it is detected early. Each year, more than 1.24 million new cases of blood cancer are reported worldwide. There are about 6,000 cancers worldwide due to this disease. The importance of cancer detection and classification has prompted researchers to evaluate Deep Convolutional Neural Networks for the purpose of classifying blood cancers. The objective of this research is to conduct an in-depth investigation of the efficacy and suitability of modern Convolutional Neural Network (CNN) architectures for the detection and classification of blood malignancies. The study focuses on investigating the potential of Deep Convolutional Neural Networks (D-CNNs), comprising not only the foundational CNN models but also those improved through transfer learning methods and incorporated into ensemble strategies, to detect diverse forms of blood cancer with a high degree of accuracy. This paper provides a comprehensive investigation into five deep learning architectures derived from CNNs. These models, namely VGG19, ResNet152v2, SEresNet152, ResNet101, and DenseNet201, integrate ensemble learning techniques with transfer learning strategies. A comparison of DenseNet201 (98.08%), VGG19 (96.94%), and SEresNet152 (90.93%) shows that DVS outperforms CNN. With transfer learning, DenseNet201 had 95.00% accuracy, VGG19 had 72.29%, and SEresNet152 had 94.16%. In the study, the ensemble DVS model achieved 98.76% accuracy. Based on our study, the ensemble DVS model is the best for detecting and classifying blood cancers.

# Introduction

According to the World Health Organization (WHO), blood cancer is one of the deadliest forms of cancer. The World Health Organization predicted that there would be about 4.5 million new cases of cancer and about 2.5 million deaths in the year 2022. When abnormal cells in the blood tissue grow out of control, they can turn into hematological cancers [1]. Blood cancer, a condition that affects both adults and children, presents one of the most challenging survival rates among diseases. The classification of different varieties of cancers may vary according to criteria including size, shape, and contrast, in addition to location, texture, and shape [2,3]. Therefore, using a CNN network to produce medical images that satisfy certain requirements like size, shape, and contrast has potential benefits for improving cancer analysis [4]. Digital images of the patient's blood, known as peripheral blood smears (PBS), are useful for the detection of blood cancers. Manual analysis takes time and is prone to errors due to the large amount of data and different blood cancer types [2]. PBS images are preferred for their non-invasiveness and ability to better highlight interior cancers [5]. Cancer sizes can be measured very precisely using peripheral blood smear (PBS) imaging. It offers insight into multiple types of blood cancers and can handle large datasets, which makes it a preferred method for blood cancer detection [5]. Despite using non-ionizing radiation, PBS produces images of a very high quality [6].

There have been recent efforts to identify and classify blood cancers by manually extracting features and detecting cancers, leading to time-consuming and inaccurate processes. For blood cancer to be extracted and diagnosed, automatic cancer detection is required [7]. Treatment of blood cancer begins with classification [3]. Automated classification is required to detect potentially fatal blood malignancies at an early stage. The adoption of three-dimensional CNNs has followed the rise of two-dimensional CNNs [8]. Another approach involves the use of D-CNN for detecting blood cancer. Blood cancer detection is a process known for its laborious nature and susceptibility to errors when utilizing medical images [5,9]. In this study, a Deep Convolutional Neural Network (D-CNN) is employed to analyze digital images for identifying and classifying blood cancers. In this study, a Deep Convolutional Neural Network (D-CNN) is used to analyze digital images in order to locate, segment, and categorize blood cancers. This study uses a diverse range of techniques, including CNNs, R-CNNs, ensemble methods, and transfer learning. A study of CNN architectures has been conducted using the VGG19, the ResNet152v2, the ResNet101, the SEresNet152, and the DenseNet201. When creating these CNNs, dimensions related to space, depth, and breadth are all taken into account. A variety of techniques are used to enhance the models, such as random search and hyperparameter optimization [10]. Previously, the study examined how deep CNNs [11] can be used to analyze images and identify blood cancers. The purpose of this study is to evaluate how effective these architectures are at identifying and classifying blood cancers. In order to classify blood cancers, CNN-based architectures are utilized. These include VGG19, ResNet152v2, ResNet101, SEresNet152, and DenseNet201.

In spite of the benefits of an ensemble model in a deep convolution neural network (D-CNN) to increase disease detection and classification accuracy, the ensemble technique has been infrequently applied to blood cancer. A method of deep learning known as ensemble learning combines a number of primary learners in a fusion strategy to enhance generalization capability. Deep learning using ensemble learning involves the integration of a large number of primary learners with fusion strategies to improve generalization [12]. As a decision-making method, ensemble learning combines multiple models to improve overall performance, encompassing classification and prediction, by exceeding the capabilities of each model. Using multiple models mitigates errors introduced by individual models, enhancing the results. In the domain of medical image analysis, specifically blood cancer research, Deep Convolutional Neural Network (D-CNN) ensemble models have drawn considerable interest. Considering D-CNNs' promising capabilities, the utilization of ensemble models for tasks like identifying and categorizing blood cancer represents an interesting area for investigation [13].

The paper is organized with a literature review, experimental setup, experimental results, discussion, and conclusion. The results are published with the experimental description due to the inclusion of three experiments in this study. The five convolutional neural networks (CNNs) used in these studies were VGG19, ResNet152v2, ResNet101, SEresNet152, and DenseNet201. A transfer learning and ensemble model were also employed. Additionally, the paper includes an analysis of its shortcomings and potential areas for future research.

# Literature review

The literature review of the study is structurally divided into two main parts. The first part reviews several CNN variations. The second part covers the general use of CNN in blood malignancies detection and classification and the specificity of accuracy in identifying particular types of blood cancer utilizing CNN-based methods. This structured approach allows us to examine both the technical aspects of CNN architecture applications and their practical implications for detecting and classifying blood cancers.

## Original CNN networks

In the context of Convolutional Neural Networks (CNNs), it integrates two functions to form a third, essentially connecting two sources of data. To extract features from the input data, CNNs employ a convolutional layer, also known as a filter or kernel, which generates a feature map [14]. The layering system of a CNN is made up of an input layer, several convolutional layers, pooling layers, a fully connected layer, and an output layer, in addition to hidden layers. CNNs are suitable for tasks that require the analysis of complex input data with spatial structure, such as image recognition and object detection.

DenseNet-201 is a 201-layer CNN that excels in image recognition. It uses a pre-trained model on a massive image database to identify objects in new images (224x224 pixels). Unlike traditional CNNs, DenseNet connects all its layers, improving feature reuse and reducing vanishing gradients (Figure 1).

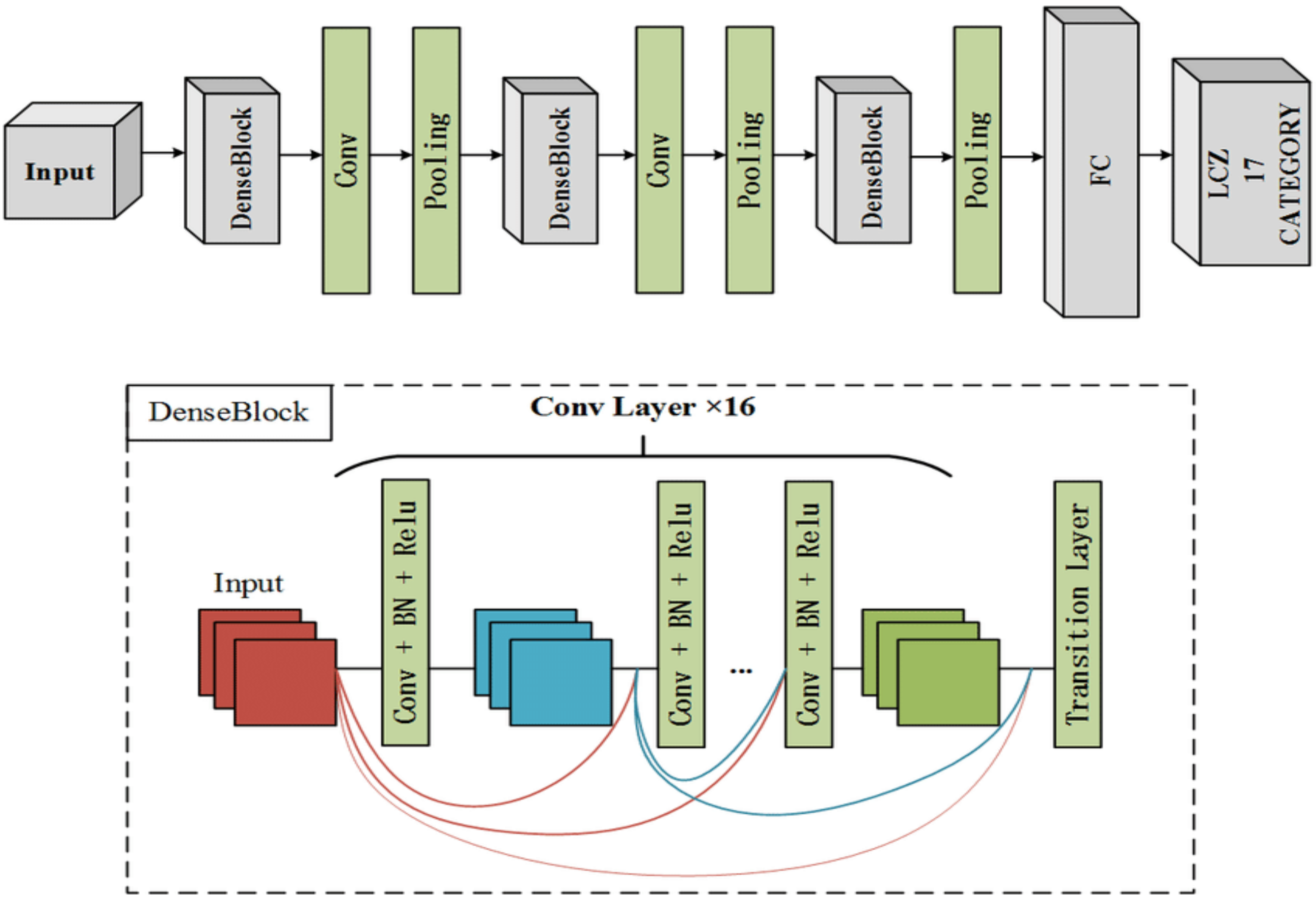

Figure 1: Schematic Representation of DenseNet-201 Architecture [27]

DenseNet's core function (D) combines convolutions, batch normalization (standardizes data) and ReLU (activates features) for efficient image processing. Image and kernel matrices are multiplied element-wise during convolution.

$$\mathrm{Densenet}(\mathrm{I}) = \mathrm{Dl}\,([\mathrm{I}, \mathrm{k1}, \mathrm{k2}, \ldots, \mathrm{k}(\mathrm{n}-1)]) \quad (1)$$

Where, I is the image and $k_1$, $k_2$, $k_{(n-1)}$ are the features of the first, second, and $k_{(n-1)}$ layers, respectively. Image spatial representation was obtained using Densenet 201 as the image feature encoder. The experiments employed 201-layer Densenet for the image model. Densenet received an image with dimensions of 224 × 224 × 3. This work's Densenet last layer output was 2208. Dense Layer shrank features from 2208 to 128. Equation (2) determines the thick layer output (a−1).

$$\mathrm{a} - 1 = \mathrm{Kd} \cdot \mathrm{Densenet}(\mathrm{I}) \quad (2)$$

Here, Kd is a kernel weights matrix having dimensions of 128 × 2208

**ResNet:** Residual Networks (ResNets) effectively address the problem of disappearing gradients in deep neural networks by using skip connections [19, 20]. This enables the utilization of far more complex models (with over 150 layers) for tasks involving image recognition, such as ResNet-50.
A network built on many stacked residual units is called ResNet50. Remaining units are used in the network's construction as its building blocks. These units consist of layers for convolution and pooling. This architecture accepts input images with a size of 224 × 224 pixels and employs 3 × 3 filters like in VGG16.

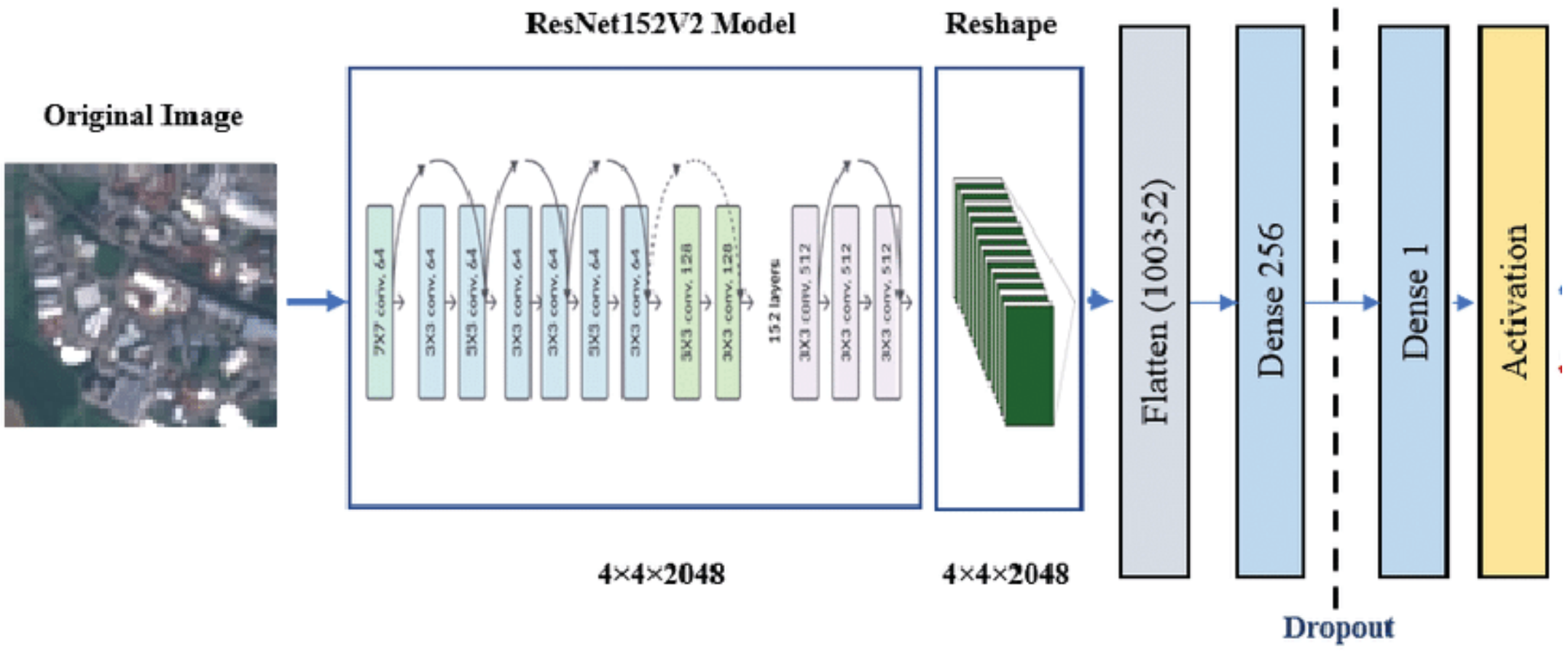

Figure 2. Schematic representation of ResNet152V2 Architecture [22].

ResNet introduced the concept of residual learning, which involves adding shortcut connections that skip one or more layers in a neural network. These shortcut connections mitigate the issue of the vanishing gradient problem and facilitate the training of far deeper networks. The Residual Block is often composed of two primary pathways: the "identity pathway" and the "shortcut pathway". The identity path denotes the initial mapping that the stacked layers want to learn, whereas the shortcut path offers an alternative route for the gradient to directly propagate across the network, bypassing each individual layer (Figure 3). The key idea is that the residual block learns a residual function, which is the difference between the output and the input.

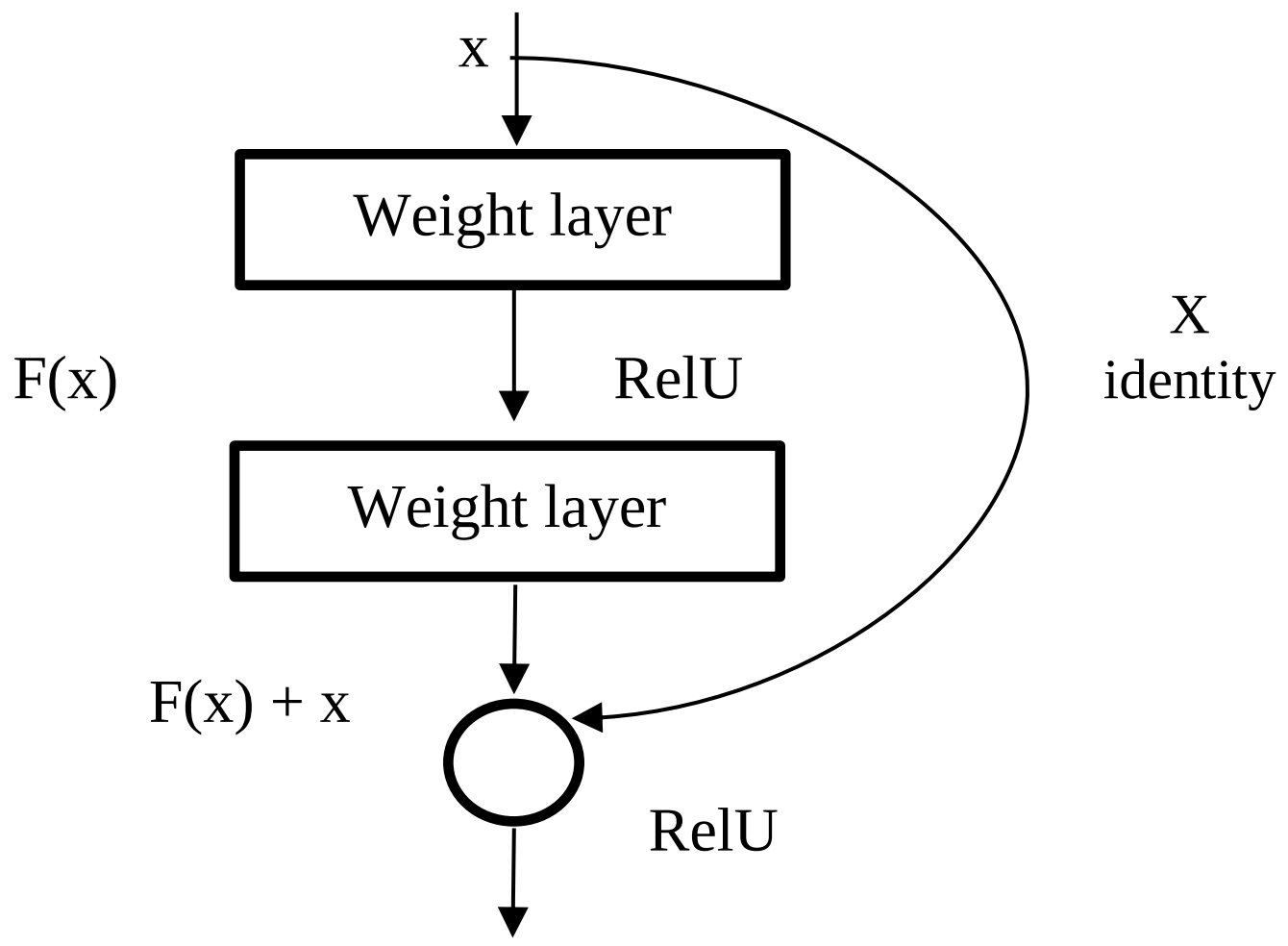

Figure 3. Residual Learning: a building block.

The identity mapping is multiplied by a linear projection Q to expand the channels of shortcut to match the residual. This allows for the input x and F(x) to be combined as input to the next layer.

$$y = F(x, \{Q_i\}) + Q_s \cdot x \quad (3)$$

**VGG19:** A variant of the VGG architecture with 19 layers facilitates the explicit and efficient transmission of spatial information among neurons within the same layer of a convolutional neural network (CNN). This approach demonstrates effectiveness particularly in scenarios where objects exhibit distinct shapes. The key advancement is a comprehensive analysis of networks with increasing depth using relatively small convolution filters while also capturing left/right and up/down features (Figure 4). There are also other 1x1 convolution filters that linearly modify the input before it is passed to the ReLU component. The convolution stride is fixed at 1 pixel to ensure that the spatial resolution is preserved after convolution [23].

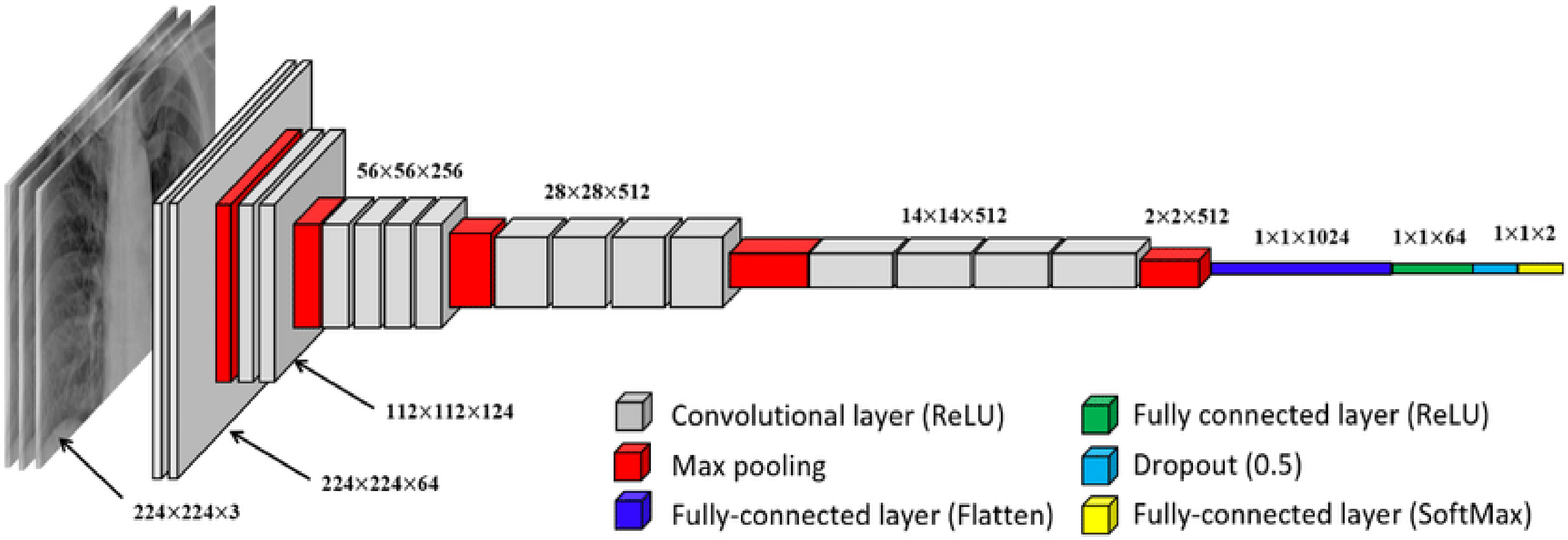

Figure 4. Schematic representation of VGG19 model Architecture [24].

The formula to calculate the output size of each convolutional layer is [(P-Q)/R] + 1.
Applying convolutions with various kernel sizes and a stride of 1.

## Case-1: When we have a kernel size of 3 x 3

- **After the First convolution**

P = 224, Q = 3, R = 1
Output shape = [(P-Q)/R] + 1 = [(224–3)/1] + 1 = 222

- **After the Second convolution**

P = 222, Q = 3, R = 1
Output shape = [(P-Q)/R] + 1 = [(222–3)/1] + 1 = 220

- **After the Third convolution**

P = 220, Q = 3, R = 1
Output shape = [(P-Q)/R] + 1 = [(220–3)/1] + 1 = 218

After three consecutive convolutions, we obtain an output size of 218 x 218 x K.

## Case-2: When we have a kernel size of 7 x 7

P = 224, Q = 7, R = 1
Output shape = [(P-Q)/R] + 1 = [(224–7)/1] + 1 = 218

A single convolution yields 218 x 218 x K. In both circumstances, the three consecutive 3x3 convolution layers have a 7x7 effective receptive field.

Table 1. Comparison of CNN architectures applied in this study

| Architecture Name | Year | Main contribution | Number of Parameters | No of layer | Reference |
|---|---|---|---|---|---|
| VGG | 2014 | - Homogenous topology<br><br>- Uses small size kernels | 138 M | 19 | Spatial Exploitation |
| Squeeze & Excitation Networks | 2017 | - Models interdependencies between feature-maps | 27.5 M | 152 | [25] |
| ResNet | 2016 | - Identity mapping-based skip connections<br>- RL: residual learning | 25.6 M<br>1.7 M | 152<br>110 | [26] |
| ResNeXt | 2017 | - Cardinality<br>- Homogeneous topology<br>- Grouped convolution | 68.1 M | 29 - 101 | [17] |
| DenseNet | 2017 | - CLF: cross-layer information flow | 25.6 M<br>25.6 M<br>15.3 M<br>15.3 M | 190<br>190<br>250<br>250 | [25] |

In addition to its numerous architectures, CNN has also led to extensive research on transfer learning and the ensemble technique. The subsequent sections will delve into these CNN techniques.

## Transfer learning

Transfer learning utilizes a pre-trained CNN, which is pretrained on a larger dataset, to expedite the training process. This method eliminates the need to train a CNN from scratch, which demands extensive labeled data and significant computational resources. Fine-tuning and function extraction are common techniques in transfer learning. Fine-tuning involves adjusting the weights of the pre-trained CNN, preserving certain levels while adapting others. Typically, early layers retain their weights, offering broad applicability, while later layers are fine-tuned to adapt to specific datasets. Alternatively, CNN can serve as a function extractor, allowing access to encoded functions from any layer to guide a chosen classifier.

## Ensemble technique

Ensemble methods in the domain of Convolutional Neural Networks (CNN) entail the use of several classifiers, which have shown superior accuracy compared to strategies that use only a single classifier. Boosting, bagging, and stacking are widely used techniques for constructing ensembles. In these approaches, the outcomes of various basic learners are combined, often using a meta-learner algorithm, to generate final predictions. To achieve these predictions, the concept of a "super learner" is utilized, which entails optimizing a loss function by utilizing the cross-validated output of the learners to ascertain suitable weights for the base learners. An ensemble seeks to overcome the shortcomings of individual models and outperform any single participating model in terms of prediction and classification by combining numerous models.

## Blood cancer literature review

Detection is the principal method utilized in the examination of blood malignancies using CNNs (Convolutional Neural Networks). The term "detection" pertains to the recognition of possible abnormalities within various tissue backgrounds. These techniques generally indicate a suspicious region on a mammography image.

### Application of blood cancer detection using CNN

Khan et al. proposed a complex method for classifying four types of white blood cells (WBCs). Their model involved multiple steps, including pre-processing, segmentation, and feature extraction. The detection model performed well on 2487 annotated blood smear images. The model distinguished the four WBC types with 98% accuracy [27].
Claro et al. introduced a convolutional neural network (CNN) architecture designed to differentiate between blood slides depicting acute lymphoblastic leukemia (ALL), acute myeloid leukemia (AML), and healthy blood slides. The study utilized 16 datasets comprising a total of 2,415 photos. The proposed model was evaluated and achieved an accuracy of 97.18% and a precision of 97.23% [28].

Abir et al. emphasized the potential of computer-aided diagnostic models in accurately detecting early leukemia, reducing the burden on physicians and enhancing diagnostic precision. The study focused on acute lymphoblastic leukemia (ALL) and employed transfer learning models, including InceptionV3, achieving an accuracy of 98.38%. Comparisons with ResNet101V2 and VGG19 were also conducted [29]. Karar et al. suggested an intelligent IoMT framework to classify acute leukemias from microscopic blood images. The framework consists of three stages: wireless digital microscopy for blood sample collection, a cloud server utilizing a GAN classifier for automatic identification of blood conditions, and medical endorsement by a hematologist. GAN-based classifier performed very well on blood cell image datasets (ALL-IDB & ASH), achieving 98.67% accuracy for ALL vs healthy and 95.5% accuracy for identifying ALL, AML, and normal cells. [30].

Sampathila et al. developed a CNN-based deep learning system to differentiate leukemic cells from normal blood cells. Their customized ALLNET model achieved impressive performance metrics, including a maximum accuracy of 95.54%, specificity of 95.81%, sensitivity of 95.91%, F1-score of 95.43%, and precision of 96%. [31]. A dual-stage CNN-based automated white blood cell differential counting system using bone marrow smear pictures was proposed by Choi et al. [32]. The model achieved a high accuracy of 97.06%, precision of 97.13%, recall of 97.1% and F-1 score. Training and testing were conducted using 2,174 patch images, and the method successfully classified images into 10 classes without the need for single cell segmentation [32]. Tusar et al. proposed utilizing deep neural networks (DNN) for automatic detection of acute lymphoblastic leukemia (ALL) blast cells in microscopic blood smear images, achieving a high 98% accuracy in detecting various ALL cell subtypes [33].

Jha et al. highlighted that leukemia (ALL) is a significant cause of global mortality, emphasizing the need for improved detection methods. They explored the potential of automated deep learning-based approaches for accurate identification. By utilizing ensemble learning on upgraded datasets, their proposed artificial neural network achieved error-free identification. With high-quality datasets, the approach achieved 100% accuracy and 96.3% accuracy even with lower-quality data [34]. Parayil et al. conducted a study to develop an automation methodology based on feature fusion. They employed fusion algorithms utilizing transfer learning methods, including VGG16 and DenseNet201 for feature extraction. The classification results were evaluated using performance indicators such as Accuracy, Precision, Recall and F1-Score. By combining feature fusion with a CNN classifier, they achieved a high accuracy of 89.75% [35].

Cheuque et al. introduced a two-stage hybrid multi-level system for categorizing lymphocytes, mononuclear monocytes, segmented neutrophils and eosinophils. It utilizes a Faster R-CNN network to differentiate between mononuclear and polymorphonuclear cells and identify the region of interest. Two parallel convolutional neural networks with the MobileNet architecture recognize the subcategories in the second stage. Monte Carlo cross-validation showed a performance metric of approximately 98.4% (Accuracy, recall, precision, and F1-score) [36]. Rastogi et al. proposed a two-step approach for accurate leukocyte classification in leukemia diagnosis. They developed a fine-tuned feature-extractor model called "LeuFeatx," based on VGG16, which effectively extracted important leukocyte characteristics from single-cell images. In binary classification on the ALL IDB2 dataset, LeuFeatx outperformed state-of-the-art techniques with a 96.15% accuracy [37].

Sneha et al. developed a deep CNN based on the Chronological Sine Cosine Algorithm (SCA) for malignancy detection in leukemia using blood smear images. Their hybrid model combines the Active Contour Model with the Fuzzy C-Means Algorithm for segmentation and analyzes segmented images for statistical and linguistic components. The proposed methodology achieved 81% accuracy in leukemia identification and performance was evaluated using precision, recall and F1 score [38]. Baig et al. proposed a model for detecting malignant leukemia cells in small blood smear images. They created a dataset of approximately 4150 images and addressed challenges such as background removal, noise reduction and image segmentation. Pre-processing techniques such as intensity adjustment and adaptive histogram equalization were applied. Binary image multiplication improved image structure and sharpness. Segmented images were obtained through operations to reduce background noise. Deep features were extracted using two trained CNN models and Canonical Correlation Analysis (CCA) fusion approach was used to combine the features. Classification algorithms including SVM, bagging ensemble, total boosts, RUSBoost and fine KNN were employed, with bagging ensemble achieving the highest accuracy of 97.04% [39].

Vogado et al. introduced LeukNet, a CNN designed for accurate leukocyte classification. Data augmentation techniques were applied to expand the training dataset and cross-validation achieved an accuracy of 98.61%. Cross-dataset validation showed that LeukNet outperformed state-of-the-art techniques with accuracy levels of 97.04%, 82.46% and 70.24% on three different datasets [40]. Vo et al. propose a method using deep learning algorithms and microscopic blood smear images to automatically detect and classify malaria and acute lymphoblastic leukemia (ALL). The method consists of three stages: segmentation using a modified UNet model, classification using a convolutional neural network and data fusion using a perceptron model. The proposed approach achieves an overall accuracy of 93%, with a 95% detection rate for ALL and a 92% detection rate for malaria. This method offers a reliable and interpretable solution for detecting abnormal leukocytes in ALL and identifying malaria-infected blood cells [41].

Table 2. Research Matrix

| Study | Image type | CNN model | Accuracy |
|---|---|---|---|
| [27] | blood smear images | convolutional neural network | 98% |
| [28] | two leukemia types on blood slide images | convolutional neural networks | 97.18% |
| [29] | microscopic blood images | transfer learning methods, including ResNet101V2 and VGG19. | 98.38% |
| [30] | microscopic blood images | The developed GAN classifier | 98.67% for binary classification and 95.5% for multi-class classification |
| [31] | microscopic images | convolutional neural network | 95.54% |
| [32] | variations in images | dual-stage convolutional neural network (CNN). | 97.06% |
| [33] | microscopic blood smears images | Deep Neural Networks | 98% |
| [34] | augmented images | Deep learning | 96.3% |
| [35] | | convoluted neural networks, DenseNet201 and VGG16 | 89.75% |
| [36] | blood smear images. | convolutional neural networks with the MobileNet | 98.4% |
| [37] | microscopic single-cell leukocyte images | VGG16 | 96.15% |
| [38] | blood smear images | The segmentation process utilizes the Mutual Information (MI) method, along with the Active Contour Model and Fuzzy C-Means Algorithm (FCM). | 81%. |
| [39] | microscopic blood smear images | Support Vector Machine (SVM), Bagging Ensemble, Total Boosts, RUSBoost, and Fine K-Nearest Neighbor (KNN) | 97.04%. |
| [40] | IoT based data | The LeukNet is a convolutional neural network (CNN) model that was inspired by the convolutional blocks of VGG-16. | 98.61%. |
| [41] | microscopic images | convolution neural network | malaria-infected blood cells with a 93% overall accuracy including the detection rate for ALL of 95% and the detection rate for malaria of 92%. |

### Knowledge gap in Blood Cancer Detection Using CNN

The literature review presents two extensively employed deep learning algorithms, namely detection and segmentation, for identifying blood cancers. The principles of merging spatial and channel information, depth and width of architecture, and multi-path information processing have received substantial attention in the field of blood cancer research [42, 43].

## Research Methodology

The tests for this study were carried out on Google CoLab using the Keras library. One of the greatest Python deep learning tools for applying machine learning methods is TensorFlow. Each model was developed by Google, trained in the cloud using a Tesla graphics processing unit, and made available through the Google Collaboratory platform (GPU). The Collaboratory framework provides up to 12GB of RAM and around 360GB of GPU in the cloud for research purposes. The original VGG19, ResNet152V2, SEresNet152, ResNext101, and DenseNet201 architectures were selected for blood cancer diagnosis in this work.

## Datasets

The images of this dataset were collected from kaggle. This dataset consisted of 3235 peripheral blood smear (PBS) images. This dataset is divided into two classes benign and malignant. The former comprises hematogenous, and the latter is the ALL group with three subtypes of malignant lymphoblasts: Malignant Early Pre-B, Malignant Pre-B, and Malignant Pro-B ALL (Table 3). The photographs were all captured with a Zeiss camera at a 100x magnification under a microscope, and they were all saved as JPG files. The sample PBS image dataset is shown in Figure 5.

Table 3. Distribution of peripheral blood smear (PBS) images used in the train, test and validation

| | No of Images | Training images | Validation images |
|---|---|---|---|
| Benign | 505 | 353 | 101 |
| [Malignant] Pro-B | 796 | 557 | 159 |
| [Malignant] Pre-B | 955 | 668 | 191 |
| [Malignant] early Pre-B | 979 | 685 | 195 |
| **Total** | **3235** | **2263** | **646** |

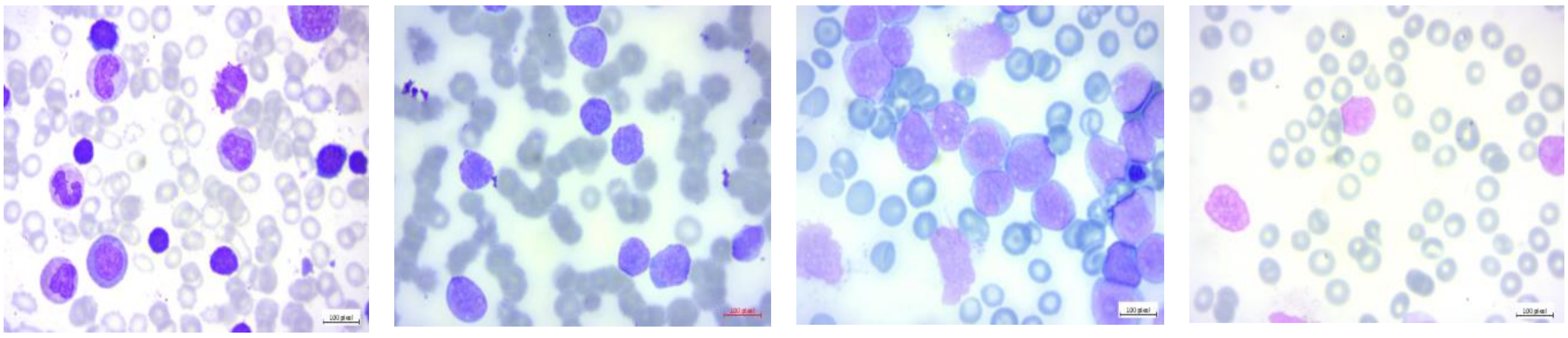

Image of Benign | Image of Malignant Early Pre-B | Image of Malignant Pre-B | Image of Malignant Pro-B

Figure 5. Example of 4 classes: Benign, Malignant Early Pre-B, Malignant Pre-B, and Malignant Pro-B.

## Process of experiments

In this step, an image enhancement is utilized. The process of expanding an existing dataset by changing the original dataset to produce more new data while simultaneously maintaining the label of the new data is known as image augmentation [45]. The objective is to raise the data set's variance while ensuring that the new data are relevant and do not bloat the dataset with irrelevant data [45]. It can enhance model generalization, make trained models more robust to unseen data, and boost model accuracy when employed in a machine learning context [45].

In order to enhance the training data, we employed data augmentation techniques with specific objectives in mind. These techniques encompassed both positional improvements, such as scaling, cropping, flipping, and rotating, as well as color improvements, including adjustments to brightness, contrast, and saturation. The data augmentation process involved various transformations, such as random rotations within the range of -15 to 15 degrees, rotations at multiples of 90 degrees, random distortion, shear transformation, vertical and horizontal flipping, skewing, and intensity transformation. Each original image was augmented in a diverse manner by randomly selecting a subset of these modifications. As a result, 10 augmented images were generated for each original image. To ensure consistency, the pixel values of both the original and augmented images used in this study were normalized by dividing them by 255. Additionally, the images were resized to a standard size compatible with all the models used in our experiment. Adjustments to the input image resolutions of all models were made to ensure uniformity throughout the study. Steps of image augmentation of PBS blood cancer images is presented in Figure 6.

Pre-processing algorithms, including Gaussian filter, LinearContrast, Median filter, and Contrast Enhancement [14], were applied to address specific objectives such as improving contrast, reducing pixel and channel noise, removing bias fields, adjusting image colors, and enhancing brightness.

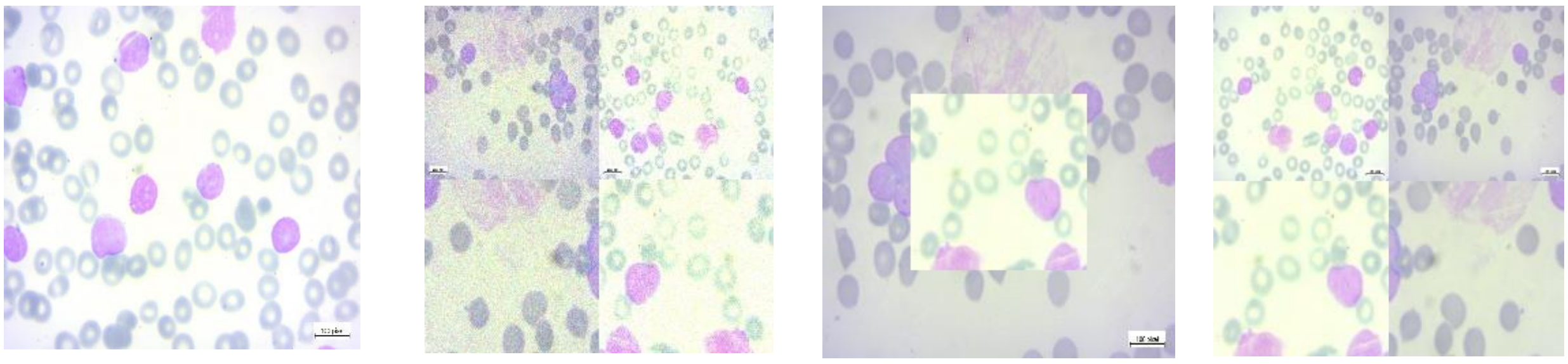

Original Image | Center augmentation | Center combined augmentation | Combined augmentation

Figure 6. Results of image augmentation.

## Training

A model was trained using the provided dataset utilizing the VGG19, ResNet152v2, DenseNet201, SEresNet152, and ResNext101 architectures, and its classification accuracy was then evaluated. Determining the optimal sizes of convolutional windows, number of layers, number of filters per layer, and other hyperparameters proved challenging in this study. Prior to deploying the CNN models in this investigation, we constructed three hyperparameter settings and performed evaluations. However, the ideal hyperparameters settings are as below:

VGG19 was trained using Early Stopping callbacks for 25 epochs; ResNet152v2, SEresNet152 and ResNext101 were trained using Early Stopping callbacks for 52 epochs and DenseNet201 was trained using Early Stopping callbacks for 18 epochs (iterations; patience = 10 iterations for all models). An Adam optimizer, Stochastic Gradient Descent (SGD) with momentum, and RMSProp were used to achieve quicker convergence (Root Mean Squared Propagation, or RMSProp, is an extension of gradient descent and the AdaGrad version of gradient descent that uses a decaying average of partial gradients in the adaptation of the step size for each parameter). The same combination was used to optimize all three models, and then they were all saved as.h5 files. While VGG19, ResNet152V2 and ResNext101 needed 43 s/epoch for model training, DenseNet201 and SEresNet152 both needed 55 s/epoch (iterations). The dataset of the experiment did not contain any major imbalances; hence standard deviation was used in this study as a model performance indicator. Categorical cross entropy was selected as the loss function for every CNN architecture because this work focuses on multi-class categorization. The final layer of the CNN topologies utilized in this work used SoftMax as the activation function, while all intermediate layers used rely. There were 60 epochs, 0.1 dropout rate, 1e-4 learning rate and batch size of 16 as the hyperparameters employed.

**Classification:** In this phase, the automatic detection of blood cancer diseases was conducted utilizing neural networks such as DenseNet121, ResNext101, ResNet152v2, ResNext101, and SEresNet152. Given its reputation as an effective classifier across numerous practical applications, the neural network was selected as the classification tool. The blood cell images were categorized into different disease classes using a softmax output layer after training the model. This model was specifically designed to identify blood cancer diseases based on the highest probability of occurrence. The experimental procedure is illustrated in Figure 7.

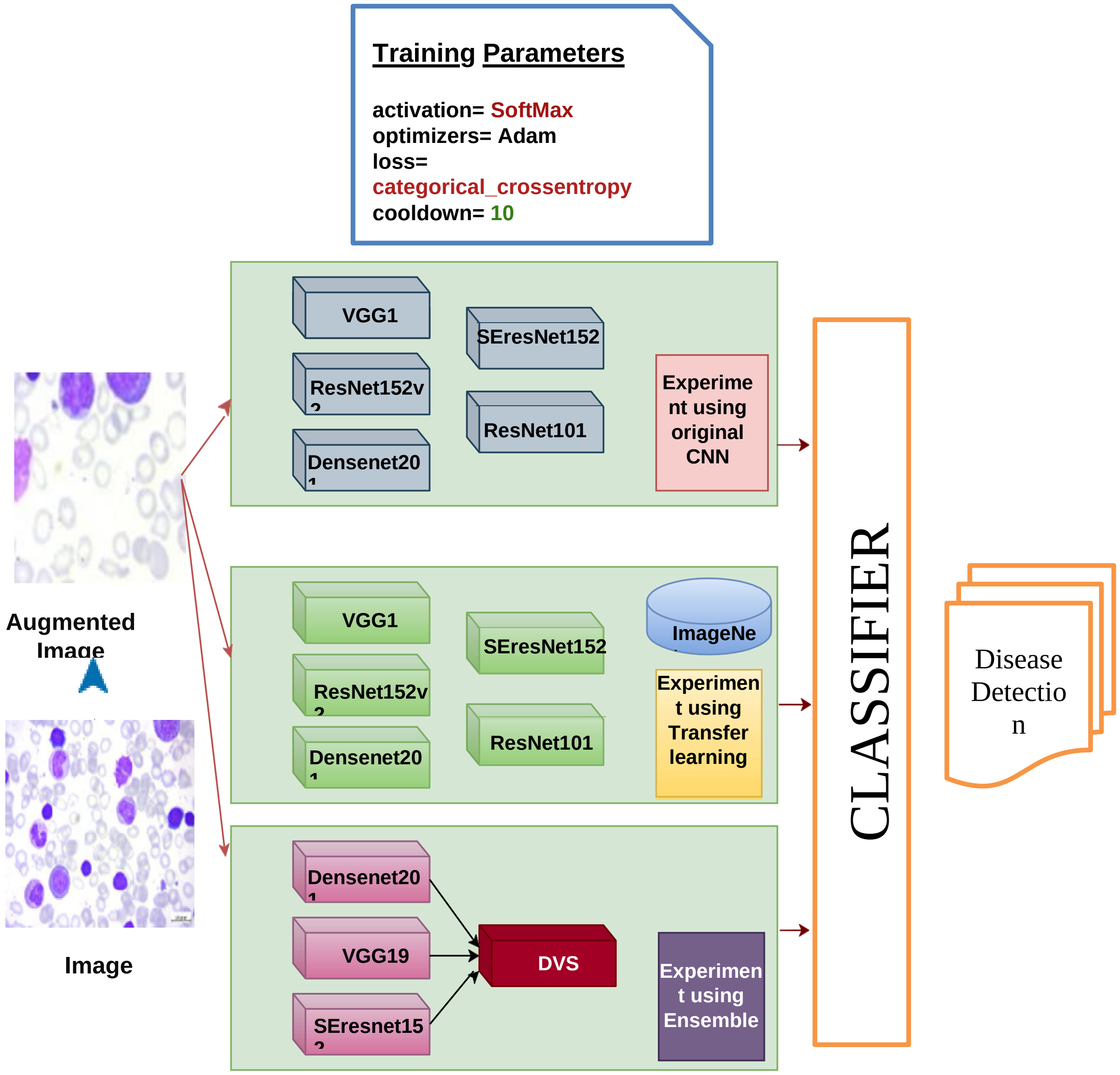

Figure 7. Diagram of the experiment.

## Results of experiments

The experiment findings are divided into three sections, focusing on the original individual network topologies, transfer learning, and ensemble methodologies. These sections aim to address the following research queries:

1. Which original CNN network exhibits higher accuracy in detecting blood cancer?
2. Does the utilization of transfer learning enhance accuracy in detecting blood cancer?
3. Does the application of ensemble techniques improve the accuracy in detecting blood cancer?

The effectiveness of such CNN base algorithms in a specific situation is measured using a variety of performance metrics for machine learning classification models. Considered are the performance indicators accuracy (AC), precision, recall, F1-score, and confusion matrix (CM). The variables true positive (TP), true negative (TN), false positive (FP), and false negative are also included in those measurements (FN).

Accuracy is one metric for measuring classification model performance. The percentage of accurate predictions the model makes is known as accuracy. Accuracy is defined in formal terms as the proportion of correctly classified images to all samples. Below is the accuracy equation:

$$Accuracy = (TP + TN)/(TP + FP + FN + TN) \quad (4)$$

Precision is the probability assigned a positive label and the proportion of those labels that are genuinely positive. It is mathematically expressed in the following equation:

$$Precision = TP/(TP + FP) \quad (5)$$

Recall is the accuracy of positive predicted instances reflecting how many were accurately labeled identified. Recall reveals the proportion of real positive cases that our model was able to properly anticipate. The F1 score is a metric used to evaluate the performance of a classification model. It considers both the precision and recall of the model to provide a single score that balances between them. It is calculated using the following equation:

$$Recall = TP/(TP + FN) \quad (6)$$

$$F1\ score = 2 \times (Precision \times Recall)/(Precision + Recall) \quad (7)$$

Specificity introduces what percentage of test results for those who do not have the disease are negative. A highly specific test is effective in excluding the majority of individuals without the disease. The equation is provided below:

$$Specificity\ = TN/(TN + FP) \quad (8)$$

The validation loss reveals how well a model fits the new data; the training loss reveals how well a model fits the training data.
The confusion matrix (CM) is a particular table format that allows the performance evaluation of an algorithm. Confusion matrices are useful because they give direct comparisons of values like TP, FP, TN, and FN. Lastly, Support is the number of actual occurrences of the class in the specified dataset. Imbalanced support in the training data may indicate structural weaknesses in the reported scores of the classifier and could indicate the need for stratified sampling or rebalancing.

The research questions of this study are addressed in the following parts based on the research questions.

## Experiment 1: Performance of the original CNN and their performance

This section presents the results of the five original individual CNN networks VGG19, ResNet152v2, DenseNet201, SEresNet152, and ResNext101. First, the models' categorization performance is shown. The final step is a discussion of the overall metrics for these models. Along with descriptions, areas for improvement in outcomes and contributing variables are gathered.

Table 4. Training and model accuracy of five original CNN architectures.

| Architecture | Training Accuracy | Model Accuracy |
|---|---|---|
| DenseNet201 | 99.65% | 98.08% |
| ResNet152v2 | 96.31% | 96.99% |
| SEresNet152 | 86.22% | 90.93% |
| ResNext101 | 85.27% | 86.41% |
| Vgg19 | 94.84% | 96.94% |

The accuracies shown in the Table 4. where the percentage of samples that could be correctly identified to all samples was calculated. In term of training accuracy, the DenseNet201 have the highest value (99.65%), compared to the architectures known as SEresNet152, and ResNext101 which have the lowest (85.22%) accuracy.  On the other hand, the model accuracy of DenseNet201 exhibits the highest percentage of 98.08% while ResNext101 has the lowest percentage of 86.41%. The remaining architectures exhibit the moderate model accuracy.

Table 5. Precision, recall, f1, and support of five (5) result of original CNN networks (based on the number of images, n= numbers)

| **VGG19** | | | | |
|---|---|---|---|---|
| | Benign | Malignant Early Pre-B | Malignant Pre-B | Malignant Pro-B |
| Precision | 97% | 94% | 99% | 98% |
| Recall | 88% | 98% | 98% | 99% |
| F1-score | 93% | 96% | 99% | 99% |
| Support (N) | 1672 | 3254 | 3198 | 2628 |
| **Resnet152 V2** | | | | |
| Precision | 96% | 95% | 98% | 99% |
| Recall | 93% | 97% | 99% | 100% |
| F1-score | 92% | 95% | 100% | 100% |
| Support (N) | 1641 | 3255 | 3196 | 2630 |
| **SEresNet152** | | | | |
| Precision | 74% | 93% | 97% | 93% |
| Recall | 81% | 85% | 95% | 99% |
| F1-score | 77% | 89% | 96% | 96% |
| Support (N) | 1671 | 3253 | 3200 | 2628 |
| **ResNext101** | | | | |
| Precision | 68% | 88% | 90% | 94% |
| Recall | 80% | 73% | 98% | 93% |
| F1-score | 74% | 80% | 94% | 94% |
| Support (N) | 1658 | 3201 | 3164 | 2601 |
| **DenseNet201** | | | | |
| Precision | 95% | 97% | 100% | 99% |
| Recall | 93% | 98% | 99% | 100% |
| F1-score | 94% | 97% | 100% | 100% |

| Support (N) | 1617 | 3255 | 3198 | 2629 |
|---|---|---|---|---|

Table 5 shows the Precision, Recall, F1-score, and Specificity acquired by the VGG19, ResNet152V2, SEresNet152, ResNext101 and DenseNet-201 models for each class. After computing precision values for each architecture on the test dataset, VGG19, DenseNet-201, and ResNet152V2 exhibit superior performance. Conversely, SEresNet152 and ResNext101 architectures demonstrate poorer performance, with the lowest identification rates.

**MN = Benign NT = Malignant Early Pre-B PT= Malignant Pre-B GL = Malignant Pro-B**

**VGG19**

| | MN | NT | PT | GL |
|---|---|---|---|---|
| MN | 1479 | 42 | 1 | 2 |
| NT | 137 | 3196 | 62 | 3 |
| PT | 9 | 6 | 3135 | 12 |
| GL | 47 | 10 | 0 | 2611 |

**ResNet152v2**

| | MN | NT | PT | GL |
|---|---|---|---|---|
| MN | 1522 | 38 | 12 | 16 |
| NT | 129 | 3211 | 4 | 66 |
| PT | 1 | 4 | 3181 | 29 |
| GL | 19 | 0 | 2 | 2518 |

**SEresNet152**

| | MN | NT | PT | GL |
|---|---|---|---|---|
| MN | 1350 | 364 | 96 | 3 |
| NT | 157 | 2781 | 63 | 0 |
| PT | 15 | 66 | 3040 | 20 |
| GL | 149 | 42 | 1 | 2605 |

**ResNext101**

| | MN | NT | PT | GL |
|---|---|---|---|---|
| MN | 1333 | 564 | 22 | 46 |
| NT | 192 | 2325 | 28 | 86 |
| PT | 34 | 283 | 3099 | 44 |
| GL | 99 | 29 | 15 | 2425 |

**DenseNet201**

| | MN | NT | PT | GL |
|---|---|---|---|---|
| MN | 1559 | 69 | 12 | 1 |
| NT | 96 | 3180 | 4 | 1 |
| PT | 1 | 6 | 3182 | 1 |
| GL | 14 | 0 | 0 | 2626 |

Figure 8. Confusion matrix of five original CNN.

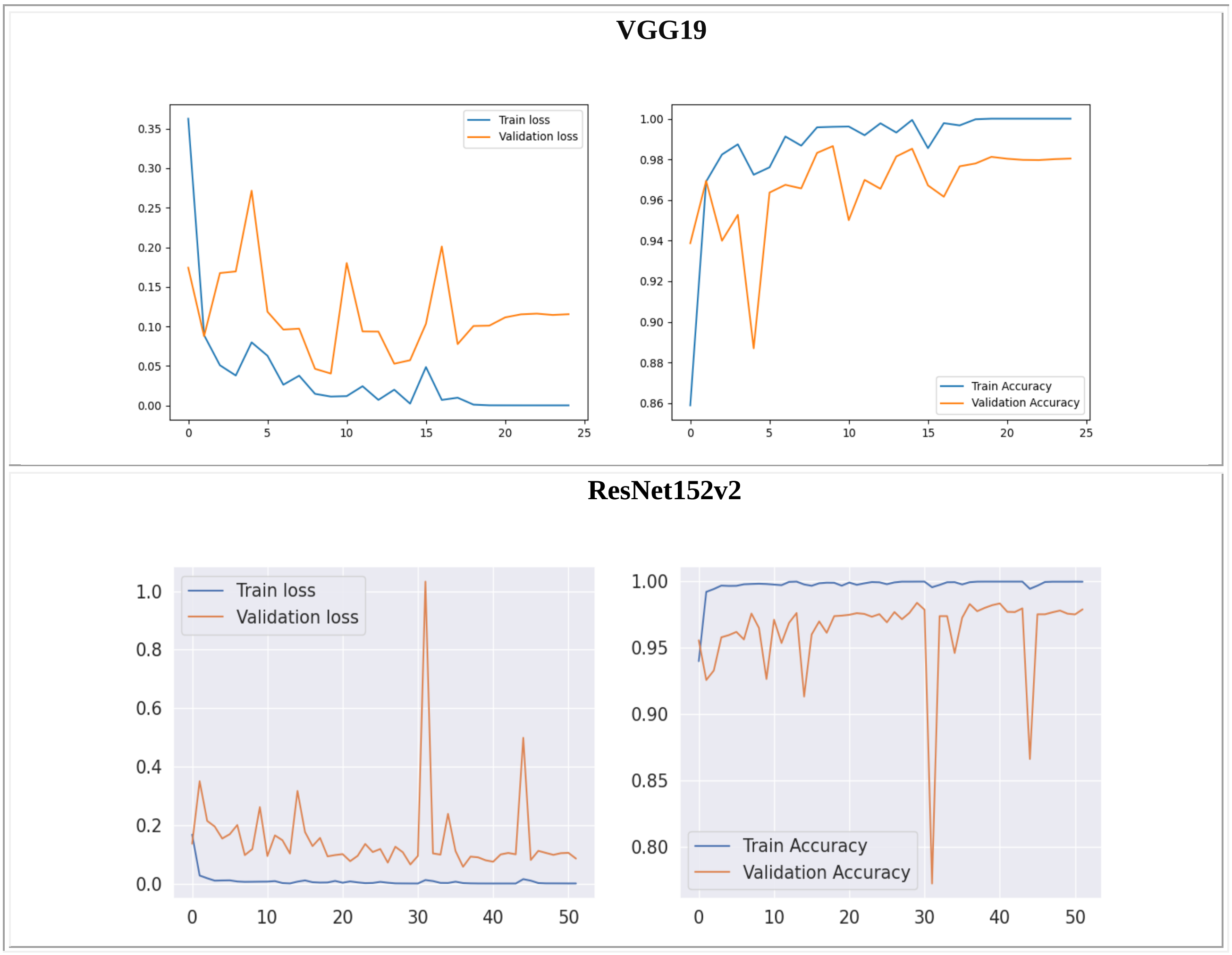

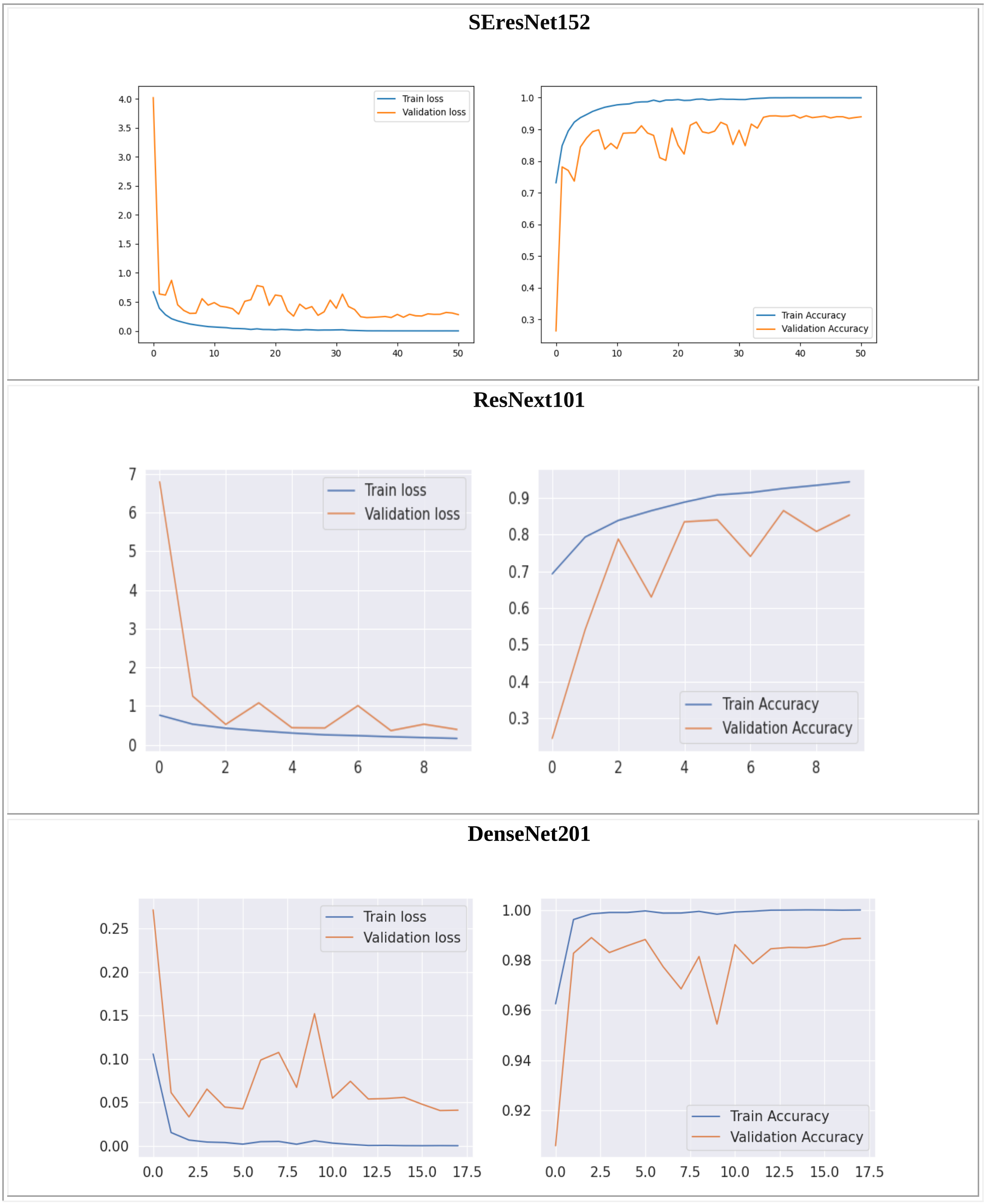

Figure 9. Loss and accuracy curve of five original CNN.

Figure 9. explains loss and accuracy curve of five original CNNs. The graphs show above illustrate the training and validation accuracy of the original model where the x-axis indicates the number of epochs, and the y-axis indicates accuracy and loss percentage. The data for all CNN models also shows that as epoch increases, both train and validation loss decrease. As the epochs increase, the loss lines show little fluctuation however, it remains steady for further epochs. Moreover, there is no over-fitting, and the training as well as the validation data are precisely split up in the figure. The loss function assists the CNN to improve the architecture. The loss value determines how accurately or poorly a model performs following each optimized iteration.

## Experiment 2: Experimental process and result of transfer learning

Five transfer learning CNN architectures' which are DenseNet-201, ResNet152v2, SecrensNet152, VGG19 and ResNext101 displays the performance of CNN models in this section. In the test sets, DenseNet-201, ResNet152v2, and SecrensNet152 models performed well, but VGG19 and ResNext101 models performed poorly.

Table 6. Training and model accuracy of Transfer learning

| **Architecture** | **Training Accuracy** | **Model Accuracy** |
|---|---|---|
| DenseNet201 | 92.57% | 95.00% |
| ResNet152v2 | 88.79% | 90.89% |
| SEresNet152 | 91.94% | 94.16% |
| Vgg19 | 75.05% | 72.29% |
| ResNext101 | 82.21% | 75.44% |

The test accuracies shown in Table 6 were calculated using the ratio of properly-identified samples to all samples. With a precision of 97 % the DenseNet201 model was the most accurate. But the DenseNet-201 network's performance dropped from its original CNN accuracy of 98.08% to 95.00% after transfer learning.

Table 7. Precision, Recall, F1, And Specificity Result of CNN Networks with Transfer Learning (N= Numbers)

| **VGG19** | | | | |
|---|---|---|---|---|
| | Benign | Malignant Early Pre-B | Malignant Pre-B | Malignant Pro-B |
| Precision | 81% | 64% | 88% | 72% |
| Recall | 25% | 78% | 94% | 79% |
| F1-score | 38% | 70% | 91% | 75% |
| Support (N) | 1672 | 3256 | 3196 | 2628 |
| **Resnet152 V2** | | | | |
| Precision | 87% | 84% | 90% | 94% |
| Recall | 66% | 90% | 98% | 90% |
| F1-score | 75% | 87% | 94% | 92% |
| Support (N) | 1670 | 3253 | 3200 | 2629 |
| **SEresNet152** | | | | |
| Precision | 94% | 87% | 93% | 96% |
| Recall | 70% | 93% | 100% | 94% |
| F1-score | 80% | 90% | 96% | 95% |
| Support (N) | 1670 | 3255 | 3199 | 2628 |
| **ResNext101** | | | | |
| Precision | 94% | 75% | 85% | 86% |
| Recall | 83% | 74% | 93% | 90% |
| F1-score | 88% | 75% | 89% | 88% |
| Support (N) | 3582 | 2946 | 3140 | 2876 |
| **DenseNet201** | | | | |
| Precision | 94% | 88% | 94% | 97% |
| Recall | 71% | 93% | 99% | 97% |
| F1-score | 80% | 91% | 96% | 97% |
| Support (N) | 1617 | 3256 | 3200 | 2625 |

The Precision, Recall, F1-score, and Specificity visualized from CNN networks incorporating transfer learning are shown in Table 7. A model is considered to be outstanding if it has high Precision, Recall, and Support. With a 64%, the trial results indicate that Vgg19 has a low precision in blood cancer.

### Experiment 3: Experimental process and result of Ensemble model

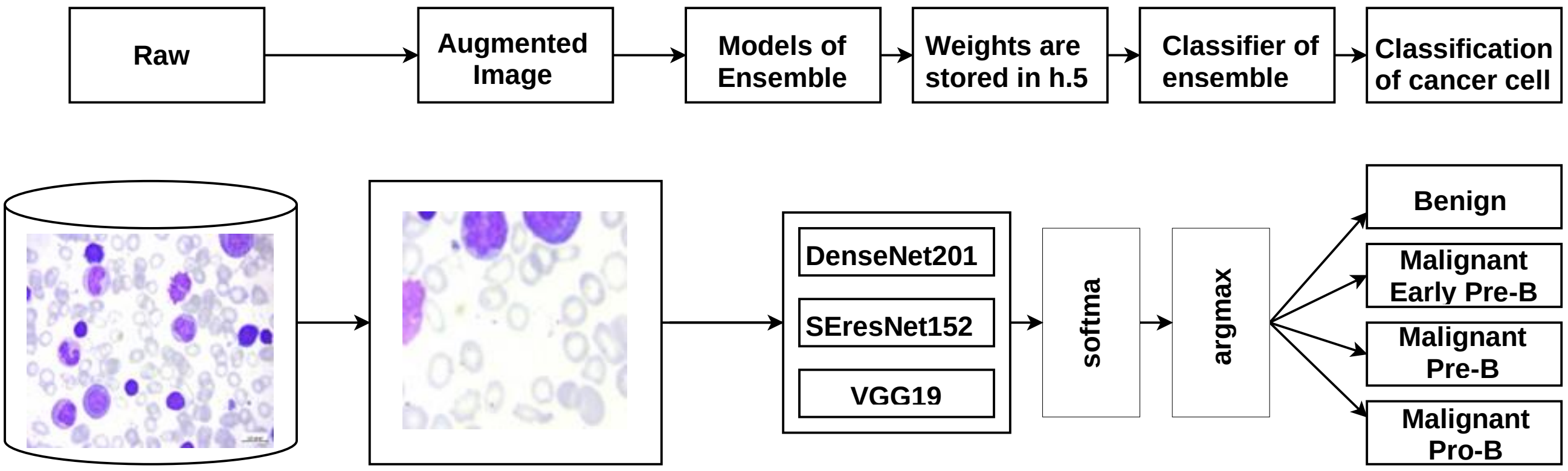

Figure 10. Block diagram of Ensemble model

In this study, three distinct original CNN models: DenseNet201, VGG19 and Serensnet152 make up the ensemble stack. To enhance the training process, we used a transfer learning approach but transfer learning performance was less expected. The output from those models was then delivered to a post-processing block with a layer that was completely related to it, a pass, and a final logits layer for classifying the image. All of the models were trained with Early Stopping (60 epochs) callbacks (patience = 10 epochs). The Adam optimizer, which combines SGD with momentum and RMSProp, was employed with the parameters learning rate to achieve faster convergence. The equivalent optimizer is applied to all three models, and the models are then saved as.h5 files. Each epoch of the DVS (DenseNet201, VGG19, SEresNet152) model training takes 68 seconds. The following graph shows how the loss function was gradually added throughout the course of the epochs for all three models: DenseNet201, VGG19, and SEresNet152.

Table 8. Training and model accuracy of Ensemble model DVS (DenseNet201, VGG19 and SEresNet152)

| **Architecture** | **Training Accuracy** | **Model Accuracy** |
|---|---|---|
| DenseNet-201, VGG19 and SEresNet152 | 97.44% | 98.76% |

Table 9. Precision, Recall, F1-Score and Support of Ensemble model DVS

| **DVS** | | | | |
|---|---|---|---|---|
| | Benign | Malignant Early Pre-B | Malignant Pre-B | Malignant Pro-B |
| precision | 98% | 92% | 95% | 99% |
| Recall | 87% | 95% | 97% | 99% |
| f1-score | 92% | 93% | 96% | 99% |
| support (N) | 1657 | 3206 | 3172 | 2589 |

According to the precision on ensemble, the algorithm achieved 99 % on Malignant Pro-B ALL, beating out transfer learning's 88% and the original CNN model's 97%. The meningioma accuracy of f1 score was the lowest of all. Table 9 displays the Precision, Recall, F1-Score and Specificity of the Ensemble model DVS when employing the ensemble.

**MN = Benign** **NT = Malignant Early Pre-B** **PT= Malignant Pre-B** **GL = Malignant Pro-B**

| | MN | NT | PT | GL |
|---|---|---|---|---|
| MN | 1436 | 10 | 1 | 17 |
| NT | 175 | 3064 | 81 | 6 |
| PT | 23 | 131 | 3088 | 2 |
| GL | 23 | 1 | 2 | 2564 |

Figure 11: Confusion matrix of Ensemble model.

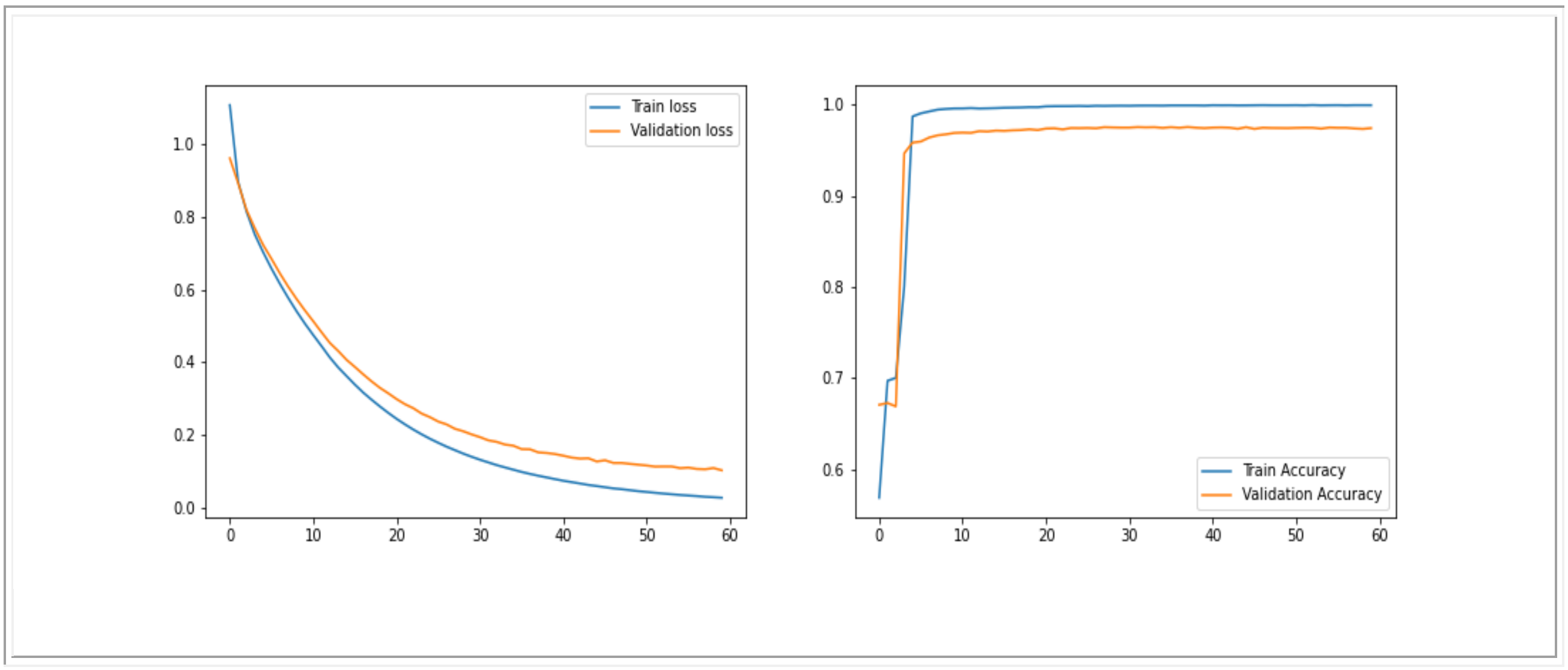

Figure 12. Loss and Accuracy curve of Ensemble model DVS.

Figure 12 illustrates the ensemble model's training accuracy and validation accuracy using data from Densenet121, VGG19, and SEresNet152. The number of epochs is shown on the x-axis, and the accuracy and loss percentages are shown on the y-axis. The Figure reveals that there is no over-fitting and that the training and validation data are divided appropriately.

# Discussions

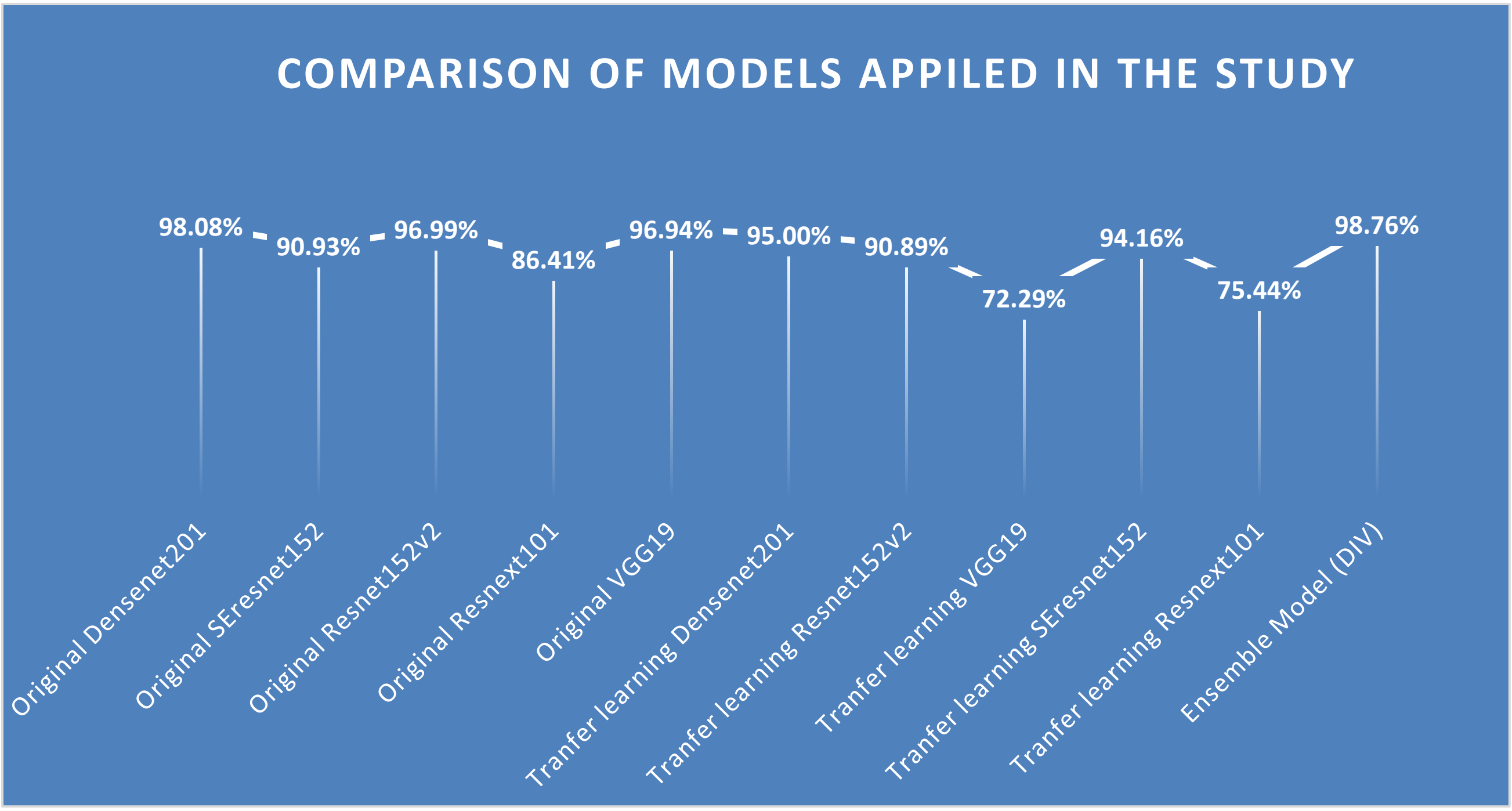

Figure 13. Accuracy comparison among individual CNN and transfer learning

In this research, we performed an in-depth investigation of the performances of the D-CNN in detecting and classifying four kinds of images of blood cancer (Benign, Malignant Early Pre-B, Malignant Pre-B, and Malignant Pro-B). Total 3235 original images were used in this research. However, image augmentation was carried and eleven image were generated using augmentation from one image. For the blood cancer image detection, there implemented three models which are the original individual CNN, transfer learning and ensemble. We compared the results of five different CNN-based models of DenseNet201, VGG19, ResNext101, ResNet152v2 and SEresnet152 by applying them to the four classes of blood cancer (see Figure 12 for accuracy). VGG19, DenseNet201, and ResNet152V2 are a few of the original individual networks that provide the best classification outcomes for diagnosing blood cancer. Additionally, DenseNet201 offers the maximum accuracy (98.08%) in the original CNN. Overall the ensemble DVS model done more performance than the original CNN.
Three models— DenseNet201, VGG19, and Seresnet201 were utilized in the ensemble technique. With a 98.76% accuracy, the ensemble model outperformed the original CNN architecture (DenseNet201, VGG19 and Seresnet201). Furthermore, it exhibits a 0.68% improvement compared to the original CNN architecture. As expected, our investigation revealed that the combination of deep learning models performed better than a single CNN architecture in terms of accuracy.

## Inference of the study

In this study, deep learning methods are investigated to detect and to segment blood cancer. The following points in this study can be concluded.

1. The classification accuracy for the same microscopic image of blood cancer with the same training set of data varies when utilizing the DenseNet201, ResNet152v2, SEresnet152, VGG19, and ResNext101 models. We obtain accuracy of 98.08%, 96.99%, 90.93%, 96.94%, and 86.41%, respectively. A CNN (convolutional neural network) with 19 layers is called VGG-19. By learning the residual representation functions rather than the signal representation directly, ResNet152 can have an extremely deep network with up to 152 layers. While each of these models address the identical image classification problem, we have compared the accuracy of the VGG19 and ResNet152 architectures. Based on the comparison at epoch 20 as a sequential method, we have come to the decision that the ResNet is the better architecture. On the other hand, SEresNet152 performs worse than the ResNet for microscopic images. As a variant of a ResNet that is constructed as a generalized feature extractor and is migrated to the target dataset. The ResNet architecture is a deep residual network expansion that substitutes a residual block that uses a "split-transform-merge" technique for the traditional residual block. A ResNext101 replicates a structural component that combines a number of transformations with the same topology. After utilizing ResNext101, compared to a ResNet152v2 microscope image, it is difficult to detect clearly that's reason accuracy ratio is poor than others. By adopting shorter connections between the layers, the DenseNet201 (Dense Convolutional Network) design aims to increase the depth of deep learning networks while also improving training efficiency. As a result, DenseNet201 was designed exclusively to increase accuracy caused by high-level neural networks' vanishing gradient, which occurs because of the distance between input and output layers.

2. In our study pin points that negative transfer which is caused from dissimilar target and source dataset. This outcome stressed the excellence of transfer learning though it has the potential to successfully train deep learning models.

3. Even if a specific CNN architecture does not perform well, an ensemble of several models still may outperform individual models. Using DenseNet201, SEresnet152, and VGG19, an ensemble model (DVS) with the best accuracy of 98.76% was proposed in this paper. Ensemble process of keeping the connections between the layers simpler using DenseNet201 and the SEresnet152 converted to a generalized feature extractor of target dataset with a deep network with up to 19 layers using VGG19.

## Contributions

A comparison study of blood cancer classification and segmentation is important to gain a full understanding of CNN performance in blood cancer research using four classes and 3235 images, we presented classification results of original, transfer learning and ensemble learning. Multiple CNN architectures, performance metrics are used for the comparative study, such as inference time, model complexity, computational complexity, and mean per class accuracy. An ensemble model was developed with an aim to increase accuracy and the study confirms that an ensemble model outperforms single CNN architecture. The ensemble model is 98.76% accuracy rate of the DVS (DenseNet201, SEresNet152 and VGG19). This shows that the DVS ensemble model has a better ability to classify blood cancer. Thus, the DVS ensemble model has better classification performance and can assist in the diagnosis of blood cancer more accurately.

## Limitations and Future Scopes

In this study, a deep learning model was developed to accurately detect and classify two distinct types of construction machinery. The network's performance, a transfer learning and ensemble model strategy.

However, it is important to acknowledge certain limitations in the current research that should be addressed in future work. Firstly, the study was constrained by the use of free-of-charge resources, specifically Google Colab. This limitation restricted the scope of experiments conducted in this study. Due to the time constraints imposed by Google Colab's limited server availability, certain aspects such as hyperparameter tuning, training the base model on databases other than ImageNet (which was used as the base database for transfer learning in this research), and the exploration of various optimizers such as Adadelta, FTRL, NAdam, and Adadelta were not performed.

Additionally, another limitation of this research lies in the use of secondary data that were publicly available, rather than collecting primary data directly from field observations or experiments. Addressing these limitations in future work will contribute to a more comprehensive and robust investigation of construction machinery detection and classification.

For the forthcoming development phase, we plan to leverage the Python Flask framework to design a user interface that empowers patients to detect and localize blood cancer diseases. The interface will not only provide detection results but also offer explanations for the detected conditions. Our objective is to propose a paradigm shift where the conventional differentiation between illness and health is replaced with a continuous spectrum of identifiable blood cell states. This approach aims to enhance the ability to predict future health risks in asymptomatic patients by identifying specific blood cell states associated with heightened risks of future diseases.

## Conclusion

Early detection and classification of blood cancer is the most necessity to correctly diagnose an affected patient. Out study suggest that detection approach using D-CNN is effective for small and dirty set of dataset. According to what we know, relatively little study has been done particularly to find blood cancer. The comparison study might be very effective for providing better blood cancer management. We evaluated the efficacy of different CNN models, including transfer learning and ensemble model in the detection of blood cancer. We discovered that ensemble DVS model of three networks DenseNet201, VGG19, and SEresNet152 delivers superior accuracy based on the accuracy.

## Declaration of competing interest

The authors declare that they have no known competing financial interests or personal relationships that could have appeared to influence the work reported in this paper.

## Data availability

The data of this research are stored in the Kaggle respiratory.

## Funding Statement

The work was not supported by any funding and neither did any of the researchers receive funds.

## References

[1] Iqbal, S., Ghani, M. U., Saba, T., & Rehman, A. (2018). Brain tumor segmentation in multi-spectral MRI using convolutional neural networks (CNN). Microscopy research and technique, 81(4), 419-427.

[2] Sun, Y., & Wang, C. (2022). A computation-efficient CNN system for high-quality brain tumor segmentation. Biomedical Signal Processing and Control, 74, 103475.

[3] Deepak, S., & Ameer, P. M. (2019). Brain tumor classification using deep CNN features via transfer learning. Computers in biology and medicine, 111, 103345.

[4] Al-qazzaz, S., Sun, X., Yang, H., Yang, Y., Xu, R., Nokes, L., & Yang, X. (2021). Image classification-based blood cancer tissue segmentation. Multimedia Tools and Applications, 80(1), 993–1008.

[5] Ayadi, W., Elhamzi, W., Charfi, I., & Atri, M. (2021). Deep CNN for blood cancer classification. Neural Processing Letters, 53(1), 671–700.

[6] Iqbal, S., Ghani, M. U., Saba, T., & Rehman, A. (2018). Brain tumor segmentation in multi-spectral MRI using convolutional neural networks (CNN). Microscopy research and technique, 81(4), 419-427.

[7] Chattopadhyay, A., & Maitra, M. (2022). MRI-based brain tumour image detection using CNN based deep learning method. Neuroscience informatics, 2(4), 100060.

[8] Khan, A. R., Khan, S., Harouni, M., Abbasi, R., Iqbal, S., & Mehmood, Z. (2021). Brain tumor segmentation using K-means clustering and deep learning with synthetic data augmentation for classification. Microscopy Research and Technique, 84(7), 1389-1399.

[9] Kibriya, H., Masood, M., Nawaz, M., & Nazir, T. (2022). Multiclass classification of brain tumors using a novel CNN architecture. Multimedia tools and applications, 81(21), 29847-29863.

[10] Hu, J., Shen, L., & Sun, G. (2018). Squeeze-and-excitation networks. In Proceedings of the IEEE conference on computer vision and pattern recognition (pp. 7132-7141).

[11] Chollet, F. (2017). Xception: Deep learning with depthwise separable convolutions. In Proceedings of the IEEE conference on computer vision and pattern recognition (pp. 1251-1258).

[12] Brunese, L., Mercaldo, F., Reginelli, A., & Santone, A. (2020). An ensemble learning approach for brain cancer detection exploiting radiomic features. Computer methods and programs in biomedicine, 185, 105134.

[13] Nanglia, S., Ahmad, M., Khan, F. A., & Jhanjhi, N. Z. (2022). An enhanced Predictive heterogeneous ensemble model for breast cancer prediction. Biomedical Signal Processing and Control, 72, 103279.

[14] Acharya, A., Muvvala, A., Gawali, S., Dhopavkar, R., Kadam, R., & Harsola, A. (2020, November). Plant Disease detection for paddy crop using Ensemble of CNNs. In 2020 IEEE International Conference for Innovation in Technology (INOCON) (pp. 1-6). IEEE.

[15] Ruba, T., Tamilselvi, R., & Beham, M. P. (2023). Brain tumor segmentation in multimodal MRI images using novel LSIS operator and deep learning. Journal of Ambient Intelligence and Humanized Computing, 14(10), 13163-13177.

[16] Huang, G., Liu, Z., Pleiss, G., Van Der Maaten, L., & Weinberger, K. Q. (2019). Convolutional networks with dense connectivity. IEEE transactions on pattern analysis and machine intelligence, 44(12), 8704-8716.

[17] Mehta, R., & Arbel, T. (2018). 3D U-Net for blood cancer segmentation. In International MICCAI Brainlesion Workshop (pp. 254–266). Springer.

[18] He, K., Zhang, X., Ren, S., & Sun, J. (2016). Identity mappings in deep residual networks. In Computer Vision–ECCV 2016: 14th European Conference, Amsterdam, The Netherlands, October 11–14, 2016, Proceedings, Part IV 14 (pp. 630-645). Springer International Publishing.

[19] Majeed, T., Rashid, R., Ali, D., & Asaad, A. (2020). Issues associated with deploying CNN transfer learning to detect COVID-19 from chest X-rays. Physical and Engineering Sciences in Medicine, 43(4), 1289-1303.

[20] Elshennawy, N. M., & Ibrahim, D. M. (2020). Deep-pneumonia framework using deep learning models based on chest X-ray images. Diagnostics, 10(9), 649.

[21] Kaldera, H. N. T. K., Gunasekara, S. R., & Dissanayake, M. B. (2019, March). Brain tumor classification and segmentation using faster R-CNN. In 2019 Advances in Science and Engineering Technology International Conferences (ASET) (pp. 1-6). IEEE.

[22] Kousalya, K., Krishnakumar, B., Aswath, A. S., Gowtham, P. S., & Vishal, S. R. (2021, November). Terrain identification and land price estimation using deep learning. In AIP Conference Proceedings (Vol. 2387, No. 1). AIP Publishing.

[23] Simonyan, K., & Zisserman, A. (2014). Very deep convolutional networks for large-scale image recognition. arXiv preprint arXiv:1409.1556.

[24] Kamil, M. Y. (2021). A deep learning framework to detect Covid-19 disease via chest X-ray and CT scan images. International Journal of Electrical & Computer Engineering (2088-8708), 11(1).

[25] Brunese, L., Mercaldo, F., Reginelli, A., & Santone, A. (2020). An ensemble learning approach for brain cancer detection exploiting radiomic features. Computer methods and programs in biomedicine, 185, 105134.

[26] Kanniappan, S., Samiayya, D., Vincent PM, D. R., Srinivasan, K., Jayakody, D. N. K., Reina, D. G., & Inoue, A. (2020). An efficient hybrid fuzzy-clustering driven 3D-modeling of magnetic resonance imagery for enhanced brain tumor diagnosis. Electronics, 9(3), 475.

[27] Khan, M. B., Islam, T., Ahmad, M., Shahrior, R., & Riya, Z. N. (2021). A CNN based deep learning approach for leukocytes classification in peripheral blood from microscopic smear blood images. In Proceedings of International Joint Conference on Advances in Computational Intelligence: IJCACI 2020 (pp. 67-76). Springer Singapore.

[28] Claro, M., Vogado, L., Veras, R., Santana, A., Tavares, J., Santos, J., & Machado, V. (2020, July). Convolution neural network models for acute leukemia diagnosis. In 2020 International Conference on Systems, Signals and Image Processing (IWSSIP) (pp. 63-68). IEEE.

[29] Abir, W. H., Uddin, M. F., Khanam, F. R., Tazin, T., Khan, M. M., Masud, M., & Aljahdali, S. (2022). Explainable AI in diagnosing and anticipating leukemia using transfer learning method. Computational Intelligence and Neuroscience, 2022.

[30] Karar, M. E., Alotaibi, B., & Alotaibi, M. (2022). Intelligent medical IoT-enabled automated microscopic image diagnosis of acute blood cancers. Sensors, 22(6), 2348.

[31] Sampathila, N., Chadaga, K., Goswami, N., Chadaga, R. P., Pandya, M., Prabhu, S., ... & Upadya, S. P. (2022, September). Customized deep learning classifier for detection of acute lymphoblastic leukemia using blood smear images. In Healthcare (Vol. 10, No. 10, p. 1812). MDPI.

[32] Choi, J. W., Ku, Y., Yoo, B. W., Kim, J. A., Lee, D. S., Chai, Y. J., ... & Kim, H. C. (2017). White blood cell differential count of maturation stages in bone marrow smear using dual-stage convolutional neural networks. PloS one, 12(12), e0189259.

[33] Tusar, M. T. H. K., & Anik, R. K. (2022). Automated detection of acute lymphoblastic leukemia subtypes from microscopic blood smear images using Deep Neural Networks. arXiv preprint arXiv:2208.08992.

[34] Jha, K. K., Das, P., & Dutta, H. S. (2022, March). Artificial neural network-based leukaemia identification and prediction using ensemble deep learning model. In 2022 International Conference on Communication, Computing and Internet of Things (IC3IoT) (pp. 1-6). IEEE.

[35] Parayil, S., & Aravinth, J. (2022, June). Transfer learning-based feature fusion of white blood cell image classification. In 2022 7th International Conference on Communication and Electronics Systems (ICCES) (pp. 1468-1474). IEEE.

[36] Cheuque, C., Querales, M., León, R., Salas, R., & Torres, R. (2022). An efficient multi-level convolutional neural network approach for white blood cells classification. Diagnostics, 12(2), 248.

[37] Rastogi, P., Khanna, K., & Singh, V. (2022). LeuFeatx: Deep learning–based feature extractor for the diagnosis of acute leukemia from microscopic images of peripheral blood smear. Computers in Biology and Medicine, 142, 105236.

[38] Sneha, D., & Alagu, S. (2021). Chronological sine cosine algorithm based deep CNN for acute lymphocytic leukemia detection. Artificial Intelligence: Advances and Application (ICAIAA 2021). https://www. researchgate. net/publication/353659892.

[39] Baig, R., Rehman, A., Almuhaimeed, A., Alzahrani, A., & Rauf, H. T. (2022). Detecting malignant leukemia cells using microscopic blood smear images: a deep learning approach. Applied Sciences, 12(13), 6317.

[40] Vogado, L., Veras, R., Aires, K., Araújo, F., Silva, R., Ponti, M., & Tavares, J. M. R. (2021). Diagnosis of leukaemia in blood slides based on a fine-tuned and highly generalisable deep learning model. Sensors, 21(9), 2989.

[41] Vo, Q. H., Le, X. H., & Le, T. H. (2022). A deep learning approach in detection of malaria and acute lymphoblastic leukemia diseases utilising blood smear microscopic images. Vietnam Journal of Science, Technology and Engineering, 64(1), 63-71.

[42] Murugesan, G. K., Nalawade, S., Ganesh, C., Wagner, B., Yu, F. F., Fei, B., ... & Maldjian, J. A. (2020). Multidimensional and multiresolution ensemble networks for brain tumor segmentation. In Brainlesion: Glioma, Multiple Sclerosis, Stroke and Traumatic Brain Injuries: 5th International Workshop, BrainLes 2019, Held in Conjunction with MICCAI 2019, Shenzhen, China, October 17, 2019, Revised Selected Papers, Part II 5 (pp. 148-157). Springer International Publishing.

[43] Murugesan, G. K., Nalawade, S., Ganesh, C., Wagner, B., Yu, F. F., Fei, B., ... & Maldjian, J. A. (2020). Multidimensional and multiresolution ensemble networks for brain tumor segmentation. In Brainlesion: Glioma, Multiple Sclerosis, Stroke and Traumatic Brain Injuries: 5th International Workshop, BrainLes 2019, Held in

Conjunction with MICCAI 2019, Shenzhen, China, October 17, 2019, Revised Selected Papers, Part II 5 (pp. 148-157). Springer International Publishing.

[44] Khan, A., Sohail, A., Zahoora, U., & Qureshi, A. S. (2020). A survey of the recent architectures of deep convolutional neural networks. Artificial intelligence review, 53, 5455-5516.

[45] Arbane, M., Benlamri, R., Brik, Y., & Djeri-Oui, M. (2021). Transfer learning for automatic blood cancer classification using MRI images. In 2020 2nd International Workshop on Human-Centric Smart Environments for Health and Well-being (IHSH) (pp. 210–214). IEEE.

[46] Rao, C. S., & Karunakara, K. (2021). A comprehensive review on brain tumor segmentation and classification of MRI images. Multimedia Tools and Applications, 80(12), 17611-17643.

[47] Rangarajan Aravind, K., & Raja, P. (2020). Automated disease classification in (Selected) agricultural crops using transfer learning. Automatika: časopis za automatiku, mjerenje, elektroniku, računarstvo i komunikacije, 61(2), 260-272.

[48] Barbedo, J. G. A. (2018). Impact of dataset size and variety on the effectiveness of deep learning and transfer learning for plant disease classification. Computers and electronics in agriculture, 153, 46-53.

[49] Paymode, A. S., & Malode, V. B. (2022). Transfer learning for multi-crop leaf disease image classification using convolutional neural network VGG. Artificial Intelligence in Agriculture, 6, 23-33.

[50] Pinto, G., Wang, Z., Roy, A., Hong, T., & Capozzoli, A. (2022). Transfer learning for smart buildings: A critical review of algorithms, applications, and future perspectives. Advances in Applied Energy, 5, 100084.